\documentclass[preprint,showpacs,preprintnumbers,amsmath,amssymb,prb]{revtex4}

\usepackage{graphicx}
\usepackage{dcolumn}
\usepackage{bm}
\usepackage{longtable}

\begin{document}

\title{Nonequilibrium sum rules for the retarded self-energy of strongly correlated electrons}

\author{V.~Turkowski} \email{turkowskiv@missouri.edu}
\affiliation{Department of Physics and Astronomy, University of
Missouri, Columbia, MO 65211}

\author{J.~K.~Freericks}
\homepage{http://www.physics.georgetown.edu/~jkf}
 \affiliation{Department of Physics, Georgetown University,
Washington, D.C. 20057}

\date{\today}

\begin{abstract}
We derive the first two moment
sum rules of the conduction electron retarded self-energy for both the
Falicov-Kimball model and the Hubbard model coupled to
an external spatially uniform and time-dependent electric field
(this derivation also extends the known nonequilibrium moment sum rules for
the Green's functions to the third moment). These
sum rules are used to further test the accuracy of nonequilibrium solutions to the
many-body problem; for example, we illustrate
how well the self-energy sum rules are satisfied for the Falicov-Kimball model in infinite
dimensions and placed in a uniform electric field turned on at time $t=0$. In general, the self-energy sum rules are satisfied to a significantly higher accuracy than the Green's functions sum rules.


\end{abstract}



\maketitle

\section{Introduction}
\label{Introduction}

   The theoretical description of nonequilibrium strongly correlated electron
   systems
   is one of the most important problems in condensed matter physics.
   This problem is not only an intellectual
   challenge, but has the potential for many practical applications.
   Systems with strong electron correlations,
   like heavy-fermion compounds, manganites, high-temperature
   superconductors and strongly correlated oxide multilayers,
   demonstrate interesting and unusual
   properties, some of which have already been applied to electronic
   and magnetic devices. Due to the expectation for strong tunability
   of such systems, they are important
   candidates to be used in modern nanoelectronics,
   like multilayered
   structures, quantum wires and dots.
Some of the properties of these materials
   can be exploited
   in spintronic and orbitronic devices, where the spin and orbital
   degrees of freedom are manipulated\cite{Tokura}.
   Since the size of modern electronic devices can be small ($\sim
   10-100$nm), the physical processes in these systems
can become strongly nonequilibrium
   because they are exposed to strong external fields, which
   are generated by moderate external potentials ($\sim 1$~V)
placed over the nanoscale structures.
   The second consequence of a small system size is that the system will have
   enhanced quantum fluctuations. This makes it difficult to study
   different properties of the system, like transport and optics,
   since we cannot use phenomenological approaches that rely on different
relaxation times (Coulomb, phonon, etc.) which are longer than
   typical timescales in the system. Recently, much progress
   has been made in experimental short pulse laser techniques, which allow one
   to study ultrafast processes in different bulk systems and
   nanostructures. These experiments also need a
   theoretical interpretation.

   Thus, it is important to have exact nonequilibrium
   solutions for correlated electron systems, which can serve as benchmarks
   for more general approximation methods. This problem is
   complicated even in the equilibrium case, due to the fact that
   one needs to treat the kinetic energy and the potential Coulomb energy terms
   in the Hamiltonian on equal footing. The simplest models
   for correlated electrons are the Hubbard model\cite{Hubbard} and
   the Falicov-Kimball model\cite{FalicovKimball} (which is a
   simplified version of the Hubbard model with localized spin-down
   electrons). The equilibrium solutions of these models are known only
   in the one-dimensional case, where an analytical Bethe ansatz
   approach\cite{LiebWu} can be used for the Hubbard model
and in the limit of infinite dimensions,
   where the dynamical mean-field theory (DMFT) can be
applied\cite{Georges,Freericks} to both models.

Similar to the equilibrium case, much progress in studying
nonequilibrium properties of correlated electron systems has been
made in both cases of low and high dimensions. Different
approaches, like perturbation theory, equation
   of motion and variational wave function methods were applied to study
   the properties of strongly correlated systems in the case
   of quantum
 dot and chain systems (see for example Refs.
 \onlinecite{Meir,Taguchi,Oka,Oka2}). Recently, a nonequilibrium
 generalization of the Bethe anzatz technique was proposed\cite{Mehta}
and simulations in one dimension with the density matrix
renormalization group have been performed\cite{schollwock}. In the
infinite- dimensional case, the nonequilibrium properties of the
Hubbard\cite{SchmidtMonien,Schmidt} and Falicov-Kimball\cite{PT}
models were studied by using second-order perturbation theory in $U$ within
DMFT. Recently, the Falicov-Kimball model was solved
exactly\cite{Nashville,spectral,denver,PRL,review,freericks_current_long} in the presence of a
homogeneous time-dependent electric field and in the case of a
sudden change in the interaction strength $U$.\cite{Eckstein} In
these papers, the nonequilibrium generalization of the DMFT
approximation was proposed, which allows one to obtain the numerical
solution of the nonequilibrium problem for the Falicov-Kimball
model. The numerical method is based on the Kadnoff-Baym-Keldysh
nonequilibrium Green's function formalism, when the nonequilibrium
Green's function is defined on the Kadanoff-Baym-Keldysh time
contour. We studied different properties of the model when a
constant electric field is switched on at a particular moment of
time. We found that Bloch oscillations of the electric current can
survive for a long time and develop beats with a period depending on the interaction strength; in addition, the Wannier-Stark peaks in the
density of states can broaden and split, when the Coulomb
interaction increases. It was also found that the Falicov-Kimball
model does not switch from one equilibrium state to another when the interaction strength is suddenly changed.

Since most solutions of strongly correlated problems are
numerical, it is important to develop tests that allow
one to check the precision of those solutions. In equilibrium, one of
the ways to check the accuracy is to calculate the
spectral moments of the Green's function\cite{white} and compare
them to exact results. Spectral moments have been used in many different contexts than just to test the numerical accuracy of numerical solutions.  Harris and Lange\cite{Harris} used spectral moments and a projection that forbids double occupancy, to determine properties about the spectral moments of the individual Hubbard bands at strong coupling.  They also determined the equilibrium Green function moments for the Falicov-Kimball model when they examined an alloy disorder Hamiltonian. Nolting\cite{Nolting} used the spectral moments to develop different strong-coupling-based approximations to the Green functions of the Hubbard model.  This approach has been extended in many different directions to look for magnetic order or to improve iterated perturbation theory in dynamical mean-field theory when away from particle-hole symmetry\cite{Geipel,Borgiel,Potthoff,Potthoff2,Eskes}.
Steven White used the exact
expressions for the zeroth and the first two spectral moments for
the Hubbard model to estimate the accuracy of a quantum
Monte Carlo solution of the
two-dimensional Hubbard model\cite{white}. Usually, only the zeroth
and the first two moments have been examined. However, as
was argued in Refs.~\onlinecite{Potthoff} and \onlinecite{Potthoff2}, it is also
important to know the third spectral moment, since it is
connected with spontaneous magnetic order in correlated systems,
and knowledge of the zeroth and the first three moments also contain
valuable information about the strongly correlated bandstructure.
The authors of these
papers have also established a relation between the zeroth and the
first moment for the self-energy with the lowest moments for
the Green's functions. This allowed them to estimate the
precision of the solution
for the self-energy at high energies. Recently, interest in the
self-energy spectral moments has been renewed, due to an
application of these results to the description of experiments on the
self-energy of high-temperature
superconductors arising from angle-resolved photoemission\cite{Kornilovitch,Randeria,Roesch}. While the retarded Green function moments we discuss here are appropriate for the full spectral function, the lesser moments (and the greater moments which can be extracted from the retarded and lesser moments) are appropriate for photoemission or inverse photoemission experiments.  The recent work in Ref.~\onlinecite{Randeria} examines the lesser moments with a further strong-coupling projection that removes doubly occupied states.  We do not examine these kinds of projections here. Instead we focus on nonequilibrium effects.

The nonequilibrium case is more
complicated than the equilibrium case. In nonequilibrium,
all Green's functions now depend on two time variables, as opposed to just the
time difference in equilibrium.
Nevertheless, exact expressions have been found\cite{spectral} for the
zeroth and the first two spectral moments of the nonequilibrium lesser and
retarded Green's functions for
Falicov-Kimball and the Hubbard models (coupled to a homogeneous and
time-dependent electric field). Surprisingly,
the retarded moments are time independent for
an arbitrary time dependence of the electric field. The moments were also used
to test the accuracy of the nonequilibrium solution to the Falicov-Kimball
model in the limit of infinite dimensions. However, as
mentioned above, it is important to also know the third
spectral moment, not only to quantitatively improve the measurement
of the accuracy of solutions, but also to extract information about the
quantum state of the system (like the renormalized band structure or the
appearance of magnetic order).

In this contribution, we generalize the results of
Ref.~\onlinecite{spectral} by deriving the third spectral moments for the
retarded and the lesser Green's functions, and deriving expressions for the
corresponding zeroth and the first spectral moments of the retarded self-energy
for the Falicov-Kimball and Hubbard models.
Surprisingly, the third-order moment of the retarded Green's function (Falicov-Kimball model)
and the zeroth (both) and first (Falicov-Kimball model) moments of the retarded self-energy remain
time-independent. We apply these results to benchmark the precision of
the DMFT solution of the Falicov-Kimball model in both the equilibrium
case (at arbitrary doping) and the nonequilibrium case (at
half-filling), when a constant electric field is switched on at a
particular moment of time.

The rest of the paper is organized as follows. The equilibrium Falicov-Kimball
and Hubbard models and their generalization to include the external
electric field are presented in Section~\ref{Formalism}. The results for
the spectral moments are presented in Sections
\ref{Green functionmoments} (Green's functions) and \ref{Sigmamoments} (self-energies).
In Section~\ref{FalicovKimball}, we give a brief description of the
nonequilibrium DMFT formalism, present equilibrium and
nonequilibrium solutions of the infinite-dimensional Falicov-Kimball
model and compare results for the moments obtained from the
numerical solutions with the exact results. Our summary
and conclusions are presented in Section~\ref{Conclusions}.

\section{\label{sec:level2} Hamiltonians for the models in equilibrium
and in a uniform field}
\label{Formalism}

The generalized equilibrium Hamiltonian for the spinless Falicov-Kimball and the
spin one-half Hubbard models can be written in the following unified form:
\begin{eqnarray}
{\mathcal H}(0)=-\sum_{ij}t_{ij}c_{i}^{\dagger}c_{j}
-\sum_{ij}t_{ij}^{f}f_{i}^{\dagger}f_{j}
-\mu\sum_{i}c_{i}^{\dagger}c_{i}
-\mu_{f}\sum_{i}f_{i}^{\dagger}f_{i}
+U\sum_{i}f_{i}^{\dagger}f_{i}c_{i}^{\dagger}c_{i}, \label{H}
\end{eqnarray}
where in the case of the Hubbard model, the operators $c_{i}$
($f_{i}$) and $f_{i}^{\dagger}$ ($f_{i}^{\dagger}$) correspond to
the spin-up (spin-down) electron annihilation and creation operators
on site $i$. In this paper, we consider the case of a hypercubic
lattice, and assume that the electrons can hop to the nearest
neighbor site. The corresponding hopping matrices are $t_{ij}=t_{ij}^{f}$
and the chemical potentials are $\mu =\mu_{f}$ for both
kinds of electrons (Zeeman splitting can be incorporated by choosing different chemical potentials, but for simplicity we keep them equal here). The last term in the Hamiltonian describes the
local Coulomb repulsion between spin-up and spin-down electrons with a
strength equal to $U$. The Hamiltonian in Eq.~(\ref{H}) also
corresponds to the spinless Falicov-Kimball model, when one sets
$t_{ij}^{f}=0$. In this case, the system consists of two kinds of
electrons: itinerant $c$-electrons and localized $f$-electrons,
which locally repel each other. In the case of the Falicov-Kimball
model, we shall also put $\mu_{f}=0$ for simplicity, since the value
of the chemical potential of the localized electrons is not
important for the spectral moments of $c$-electrons, which we
evaluate below.

The electric field ${\bf E}({\bf r}, t)$ can be introduced into the
Hamiltonian by means of the Peierls substitution for
the hopping matrices \cite{Jauho}:
\begin{eqnarray}
t_{ij}\rightarrow t_{ij}\exp\left[ -\frac{ie}{\hbar c}\int_{{\bf
R}_{i}}^{{\bf R}_{j}}{\bf A}({\bf r}, t)d{\bf r} \right]
 , \label{Peierls1}
\end{eqnarray}
\begin{eqnarray}
t_{ij}^{f}\rightarrow t_{ij}^{f}\exp\left[ -\frac{ie}{\hbar c}
\int_{{\bf R}_{i}}^{{\bf R}_{j}}{\bf A}({\bf r}, t)d{\bf r} \right]
, \label{Peierls2}
\end{eqnarray}
where the electric vector potential ${\bf A}({\bf r}, t)$ is
connected to the electric field in the following way:
\begin{equation}
{\bf E}({\bf r}, t)=-\frac{1}{c}\frac{\partial {\bf A}({\bf r}, t)}
{\partial t} \label{Electricfield}
\end{equation}
and the scalar potential vanishes.
This choice of the electromagnetic potential, when the scalar
potential is set equal to zero, corresponds to the Hamiltonian gauge. For
simplicity, we also assume that the electric field is spatially uniform
and it lies along the direction of the elementary cell diagonal:
\begin{equation}
{\bf A}({\bf r},t)=A(t)(1,1,...,1). \label{A}
\end{equation}
Neglecting the spatial dependence of the vector potential, assumes
that we neglect the magnetic field effects in the system [since the
magnetic field is ${\bf H}({\bf r}, t)={\bf \nabla}\times {\bf
A}({\bf r}, t)$], assuming that the electric field is smooth enough
in time, that the transient magnetic field can be neglected. This
can take place in nanostructures, when an applied external potential
produces an almost homogeneous electric field due to the small size
of the system (see also the discussion in
Ref.~\onlinecite{spectral}).

The Hamiltonian (in the Schr\"odinger picture), which describes the electron system coupled to an
external spatially independent electric field, has a rather simple
form in the momentum representation (the creation and annihilation operators now create or annihilate electrons with definite momentum):
\begin{eqnarray}
\mathcal{H}({\bf A}) =&~&\sum_{{\bf k}}\left[\epsilon \left({\bf
k}-\frac{e{\bf A}(t)}{\hbar c}\right)
-\mu\right] c_{{\bf k}}^{\dagger}c_{{\bf k}}\nonumber\\
&+&\sum_{{\bf k}} \left[\epsilon^{f}\left({\bf k}-\frac{e{\bf
A}(t)}{\hbar c}\right) -\mu_{f}\right] f_{{\bf k}}^{\dagger}f_{{\bf
k}} +U\sum_{{\bf p},{\bf k},{\bf q}} f_{{\bf p}+{\bf
q}}^{\dagger}c_{{\bf k}-{\bf q}}^{\dagger} c_{{\bf k}}f_{{\bf p}},
\label{Hk}
\end{eqnarray}
where the free electron bandstructures are:
\begin{eqnarray}
\epsilon\left({\bf k}-\frac{e{\bf A}(t)}{\hbar c}\right)
=\epsilon^{f}\left({\bf k}-\frac{e{\bf A}(t)}{\hbar c}\right)
\label{Ek} =-2t \sum_{j=1}^{d}\cos\left[a\left( {\bf
k}_{j}-\frac{e{\bf A}_{j}(t)}{\hbar c} \right)\right] ,
\label{freespectra}
\end{eqnarray}
$d$ is dimensionality of the system and $t$ is the corresponding hopping
parameter. In the case of the Falicov-Kimball model, one has to put
$\epsilon^{f}\left({\bf k}-\frac{e{\bf A}(t)}{\hbar c}\right)
-\mu_{f}=0$ in Eq.~(\ref{Hk}).

\section{Spectral moments for the Green's functions}
\label{Green functionmoments}

 In the case of nonequilibrium, there are two independent
Green functions, which describe the properties of a many-body system.
We use the retarded
\begin{equation}
G_{\bf k}^{R}(t_{1},t_{2}) =-i\theta (t_{1}-t_{2})\left\langle
\left\{ c_{{\bf k}}(t_{1}),c_{{\bf
k}}^{\dagger}(t_{2})\right\}\right\rangle \label{GR}
\end{equation}
 and the lesser
\begin{equation}
G_{\bf k}^{<}(t_{1},t_{2}) =i\left\langle c_{{\bf
k}}^{\dagger}(t_{2})c_{{\bf k}}(t_{1})\right\rangle \label{Gl}
\end{equation}
Green functions as the basis functions. The fermion operators on the
right hand side of Eqs.~(\ref{GR}) and (\ref{Gl}) are in the
Heisenberg representation and the averaging operation $\left\langle
... \right\rangle$ is performed with respect to the equilibrium
Hamiltonian (corresponding to the initial conditions prior to the field being turned on). It is convenient to use the Green functions in Eqs.~(\ref{GR})
and (\ref{Gl}), since they have important physical interpretations.
Namely, the poles of the retarded Green function define the energy
levels of the system (and thereby determine the many-body density of states), and the equal time lesser Green function
describes the occupation of these levels (and hence determine the distribution function). In equilibrium, only one
of these functions is independent, since they are connected by a
simple relation depending on the Fermi-Dirac distribution.

In order to calculate moments of the spectral functions at different
values of time, it is convenient to introduce Wigner's time
variables for the Green functions in Eqs.~(\ref{GR}) and (\ref{Gl}): the average
time $T=(t_{1}+t_{2})/2$ and the relative time $t=t_{1}-t_{2}$. The
frequency dependence of a Green function can be calculated by Fourier
transforming the Green function with respect to the relative time coordinate,
and the time evolution of the function is then described by the average
time coordinate. In other words, the average time
coordinate is associated with the physical time in the system.
The spectral function for the retarded and the lesser Green functions
can then be defined in the following way:
\begin{eqnarray}
A_{{\bf k}}^{R,<}(T,\omega) =
\frac{\eta}{\pi} {\rm Im}\int_{-\infty}^{\infty}dt e^{i\omega t}G_{{\bf k}}^{R,<}(T,t),
\label{AR<}
\end{eqnarray}
where we have introduced a prefactor $\eta$, equal to $-1$ for the
retarded Green function and $1$ for the lesser Green function in order
to have positive
zeroth moments for both retarded and lesser Green functions (see below). The
$n$th spectral moments that correspond to the spectral functions in
Eq.~(\ref{AR<}) are defined to be
\begin{eqnarray}
\mu_{n}^{R,<}({\bf k},T)&=&\int_{-\infty}^{\infty}d\omega \omega^{n}
A_{{\bf k}}^{R,<}(T,\omega). \label{munR<}
\end{eqnarray}
It is not difficult to show from Eqs.~(\ref{AR<}) and (\ref{munR<})
that there exist the following relations that connect the moments
with the corresponding Green functions:
\begin{eqnarray}
\mu_{n}^{R,<}({\bf k},T)= \frac{\eta}{\pi}{\rm Im}
\int_{-\infty}^{\infty}d\omega  \int_{-\infty}^{\infty}dt
e^{i\omega t} i^{n}\frac{\partial^{n}}{\partial t^{n}} G_{{\bf
k}}^{R,<}(T,t) \label{munR<2}
\end{eqnarray}
and
\begin{equation}
\mu_{n}^{R,<}({\bf k},T)= 2\frac{\eta}{\pi}{\rm Im} \left[
i^{n}\frac{\partial^{n}}{\partial t^{n}} G_{{\bf k}}^{R,<}(T,t)
\right]_{t=0^{+}}  \label{munR<5}
\end{equation}
(for details, see Ref.~\onlinecite{spectral}). It is more convenient to
use the expression in Eq.~(\ref{munR<2}) for the retarded Green function,
and in Eq.~(\ref{munR<5}) for the lesser Green function. The time
derivatives with respect to the operators of the Green functions in
Eqs.~(\ref{GR})
and (\ref{Gl}) can be expressed by taking commutators of the
corresponding fermion operators with the Hamiltonian in the Heisenberg picture (the terms
proportional to the time derivatives with respect to the theta function
in the case of the retarded Green function do not contribute to the moments).
This leads to the following expressions, which connect the zeroth
and the first three spectral moments with specific correlation
functions:
\begin{eqnarray}
\mu_{0}^{R}({\bf k},T)&=&\langle \{c_{k}(T),c_{k}^{\dagger}(T)\} \rangle , \label{mu0R} \\
\mu_{1}^{R}({\bf k},T)&=&\frac{1}{2}\left[ \langle
\{L^{1}c_{k}(T),c_{k}^{\dagger}(T)\} \rangle -\langle
\{c_{k}(T),L^{1}c_{k}^{\dagger}(T)\} \rangle \right], \label{mu1R} \\
\mu_{2}^{R}({\bf k},T)&=&\frac{1}{4}\left[ \langle
\{L^{2}c_{k}(T),c_{k}^{\dagger}(T)\} \rangle -2\langle
\{L^{1}c_{k}(T),L^{1}c_{k}^{\dagger}(T)\} \rangle +\langle
\{c_{k}(T),L^{2}c_{k}^{\dagger}(T)\} \rangle \right], \label{mu2R} \\
\mu_{3}^{R}({\bf k},T)&=&\frac{1}{8}\left[ \langle
\{L^{3}c_{k}(T),c_{k}^{\dagger}(T)\} \rangle -3\langle
\{L^{2}c_{k}(T),L^{1}c_{k}^{\dagger}(T)\} \rangle +3\langle
\{L^{1}c_{k}(T),L^{2}c_{k}^{\dagger}(T)\} \rangle \right.
\nonumber \\
&~&\left. -\langle \{c_{k}(T),L^{3}c_{k}^{\dagger}(T)\} \rangle
\right], \label{mu3R}
\end{eqnarray}
\begin{eqnarray}
\mu_{0}^{<}({\bf k},T)&=&2\langle c_{k}^{\dagger}(T)c_{k}(T) \rangle , \label{mu0<} \\
\mu_{1}^{<}({\bf k},T)&=&\langle c_{k}^{\dagger}(T) L^{1}c_{k}(T)
\rangle -\langle
\left( L^{1}c_{k}^{\dagger}(T)\right) c_{k}(T) \rangle , \label{mu1<} \\
\mu_{2}^{<}({\bf k},T)&=&\frac{1}{2}\left[ \langle
c_{k}^{\dagger}(T) L^{2}c_{k}(T) \rangle -2\langle \left(
L^{1}c_{k}^{\dagger}(T)\right) \left( L^{1}c_{k}(T)\right) \rangle
+\langle
\left( L^{2}c_{k}^{\dagger}(T)\right)c_{k}(T)\rangle \right], \label{mu2<} \\
\mu_{3}^{<}({\bf k},T)&=&\frac{1}{4}\left[ \langle
c_{k}^{\dagger}(T) L^{3}c_{k}(T) \rangle -3\langle \left(
L^{1}c_{k}^{\dagger}(T)\right) \left( L^{2}c_{k}(T)\right) \rangle
+3\langle \left( L^{2}c_{k}^{\dagger}(T)\right)\left(
L^{1}c_{k}(T)\right) \rangle \right.
\nonumber \\
&~&\left. -\langle \left( L^{3}c_{k}^{\dagger}(T)\right)
c_{k}(T)\rangle \right], \label{mu3<}
\end{eqnarray}
where $L^{n}O=[...[[O,\mathcal{H}_H(T)],\mathcal{H}_H(T)]...\mathcal{H}_H(T)]$ is the multiple commutation operator with respect to the Hamiltonian (in the Heisenberg picture), performed $n$ times; the operator $\mathcal{H}_H(T)$ is given by Eq.~(\ref{Hk}) with all fermionic operators replaced by the Heisenberg-picture operators evaluated at time $T$. The commutation relations can be evaluated directly because two fermionic operators at equal times (within the Heisenberg picture) satisfy canonical commutation relations.

Evaluating the commutation and anticommutation operations in
Eqs.~(\ref{mu0R})-(\ref{mu3R}) results in the following expressions
for the retarded moments:
\begin{eqnarray}
\mu_{0}^{R}({\bf k},T)&=&1,\\
\mu_{1}^{R}({\bf k},T)&=&[\varepsilon(k-eA(T))-\mu
]+Un_{f},\\
\mu_{2}^{R}({\bf k},T)&=&[\varepsilon(k-eA(T))-\mu
]^{2}+2U[\varepsilon(k-eA(T))-\mu ]n_{f}+U^{2}n_{f},
\label{mukRresults02}
\end{eqnarray}
\begin{eqnarray}
\mu_{3}^{R}({\bf k},T)&=&[\varepsilon(k-eA(T))-\mu
]^{3}+3U[\varepsilon(k-eA(T))-\mu ]^{2}n_{f}
+3U^{2}[\varepsilon(k-eA(T))-\mu ]n_{f}
\nonumber \\
&+&U^{2}\sum_{p,q}[\varepsilon^{f}(p+q-eA(T))-2\varepsilon^{f}(p-eA(T))
+\varepsilon^{f}(p-q-eA(T))] \langle f_{p}^{\dagger}f_{p}\rangle (T)
\nonumber \\
&-&U^{2}\sum_{p,q,q'}[\varepsilon^{f}(p+q-eA(T))-\varepsilon^{f}(p+q+q'-eA(T))
\nonumber \\
&-&\varepsilon^{f}(p-eA(T)) +\varepsilon^{f}(p+q'-eA(T))] \langle
f_{p+q+q'}^{\dagger}f_{p}c_{k-q}^{\dagger}c_{k+q'}\rangle (T)
\nonumber \\
&+&U^{2}\sum_{p,p',q}[\varepsilon (k+q-eA(T))-\varepsilon (k-eA(T))
+\varepsilon^{f}(p'-eA(T)) -\varepsilon^{f}(p'-q-eA(T))
\nonumber \\
&+&2\varepsilon^{f} (p-eA(T))-2\varepsilon^{f}(p+q-eA(T))] \langle
f_{p'-q}^{\dagger}f_{p'}f_{p+q}^{\dagger}f_{p}\rangle (T)
+U^{3}n_{f}. \label{mukRresult3}
\end{eqnarray}
Summing over momentum yields the following local moments:
\begin{eqnarray}
\mu_{0}^{R}(T)&=&1, \label{mulocRresult0}\\
\mu_{1}^{R}(T)&=&-\mu+Un_{f},\label{mulocRresult1}\\
\mu_{2}^{R}(T)&=&\frac{t^{*2}}{2}+\mu^{2} -2U\mu n_{f}+U^{2}n_{f},
\label{mulocRresult2}
\end{eqnarray}
\begin{eqnarray}
\mu_{3}^{R}(T)&=&-\frac{3t^{*2}}{2}(\mu -Un_{f})+3U\mu n_{f}(\mu
-Un_{f})+3U^{2}\mu n_{f}(n_{f}-1)+U^{3}n_{f}-\mu^{3} \nonumber \\
&~&+2U^{2}\sum_{ij}{\tilde t}_{ij}^{f}\langle
f_{i}^{\dagger}f_{j}\rangle -2U^{2}\sum_{ij}{\tilde
t}_{ij}^{f}\left(\langle
f_{i}^{\dagger}f_{j}c_{j}^{\dagger}c_{j}\rangle +\langle
f_{i}^{\dagger}f_{j}c_{i}^{\dagger}c_{i}\rangle \right) \nonumber \\
&~&+U^{2}\sum_{ij}{\tilde t}_{ij}^{f}\left( -\langle
f_{i}^{\dagger}f_{j}f_{i}^{\dagger}f_{i}\rangle +\langle
f_{i}^{\dagger}f_{j}f_{j}^{\dagger}f_{j}\rangle -2\langle
f_{i}^{\dagger}f_{i}f_{i}^{\dagger}f_{j}\rangle +2\langle
f_{j}^{\dagger}f_{j}f_{i}^{\dagger}f_{j}\rangle\right) .
 \label{localmuRresult3}
\end{eqnarray}
where ${\tilde t}_{ij}^{f}=t_{ij}^{f}\exp[-i\int_{R_{j}}^{R_{i}}{\bf
A}({\bf r},t)d{\bf r}]$. In these equations, we have assumed we are on the infinite-dimensional hypercubic lattice, and have evaluated the second moment of the hopping matrix explicitly; the generalization to finite dimensions is simple to complete (see the erratum of Ref.~\onlinecite{spectral}).

As follows from Eqs.~(\ref{mulocRresult0})-(\ref{mulocRresult2}),
the zeroth and the first two retarded moments remain time independent
even in the case of an arbitrary external time-dependent field. The
third local moment [in Eq.~(\ref{localmuRresult3})] is time-independent
for the case of the Falicov-Kimball model ($\tilde t^f=0$).
In the case of the Hubbard
model, its expression is complex and we cannot immediately tell whether they
are time dependent (but they most likely are). The last two terms in
Eq.~(\ref{localmuRresult3}) are defined by electron correlations and
they define the shape of the spectral functions of the lower and
upper Hubbard bands, the redistribution of the spectral weights
between the bands and a shift of their centers of
gravity\cite{Potthoff,Potthoff2}. It is difficult to obtain analytical
expressions for these terms.

In a similar way, one can obtain expressions for the lesser moments
from Eqs.~(\ref{mu0<})-(\ref{mu3<}):
\begin{eqnarray}
\mu_{0}^{<}({\bf k},T)&=&2\langle n_{k}(T)\rangle\\
\mu_{1}^{<}({\bf k},T)&=&2[\varepsilon(k-eA(T))-\mu ]\langle
n_{k}(T)\rangle \nonumber \\
&~&+ U\sum_{p,q}\left[ \langle
c_{k}^{\dagger}c_{k+q}f_{p+q}^{\dagger}f_{p}\rangle (T) + \langle
c_{k-q}^{\dagger}c_{k}f_{p+q}^{\dagger}f_{p}\rangle (T)\right] \\
\mu_{2}^{<}({\bf k},T)&=&2[\varepsilon(k-eA(T))-\mu ]^{2}\langle
n_{k}(T)\rangle \nonumber \\
&~&+\frac{3}{2}U[\varepsilon(k-eA(T))-\mu ] \sum_{p,q}[\langle
f_{p+q}^{\dagger}f_{p} c_{k-q}^{\dagger}c_{k}\rangle (T) +\langle
f_{p+q}^{\dagger}f_{p} c_{k}^{\dagger}c_{k+q}\rangle ] \nonumber \\
&~&+\frac{1}{2}U \sum_{p,q}[\varepsilon(k-q-eA(T))-\mu ] \langle
f_{p+q}^{\dagger}f_{p} c_{k-q}^{\dagger}c_{k}\rangle (T) \nonumber
\\
&~&+\frac{1}{2}U \sum_{p,q}[\varepsilon(k+q-eA(T))-\mu ] \langle
f_{p+q}^{\dagger}f_{p} c_{k}^{\dagger}c_{k+q}\rangle (T)
\nonumber \\
&~&-\frac{1}{2}U
\sum_{p,q}[\varepsilon^{f}(p+q-eA(T))-\varepsilon^{f}(p-eA(T)) ]
[\langle f_{p+q}^{\dagger}f_{p} c_{k}^{\dagger}c_{k+q}\rangle (T)
 -\langle f_{p+q}^{\dagger}f_{p} c_{k-q}^{\dagger}c_{k}\rangle (T)]
\nonumber \\
&~&+\frac{1}{2}U^{2}\sum_{p,q,P,Q} \left[ \langle
f_{p+q}^{\dagger}f_{p}f_{P+Q}^{\dagger}f_{P}
c_{k-q-Q}^{\dagger}c_{k}\rangle(T) +2\langle
f_{p+q}^{\dagger}f_{p}f_{P+Q}^{\dagger}f_{P}
c_{k-q}^{\dagger}c_{k+Q}\rangle(T) \right. \nonumber \\
&~&\left. +\langle f_{p+q}^{\dagger}f_{p}f_{P+Q}^{\dagger}f_{P}
c_{k}^{\dagger}c_{k+Q+q}\rangle(T) \right] , \label{muk<results02}
\end{eqnarray}
\begin{eqnarray}
\mu_{3}^{<}({\bf k},T)&=&2[\varepsilon(k-eA(T))-\mu ]^{3}\langle
c^{\dagger}_{k}c_{k}\rangle (T) +2U(\varepsilon(k-eA(T))-\mu
)^{2}\sum_{p,q}\langle
c_{k}^{\dagger}c_{k+q}f_{p+q}^{\dagger}f_{p}\rangle (T)
\nonumber \\
&+&2U\sum_{p',p}[\varepsilon (k-eA(T))+\varepsilon (p-eA(T))-2\mu
\nonumber \\
&+&\varepsilon^{f}(p'-eA(T)) -\varepsilon^{f}(p'+p-k-eA(T))]
(\varepsilon (p-eA(T)) -\mu )
 \langle
c_{k}^{\dagger}c_{p}f_{p'+p-k}^{\dagger}f_{p'}\rangle (T)
\nonumber \\
&+&2U\sum_{q',p}[\varepsilon (k-eA(T))+\varepsilon (k+q'-eA(T))-2\mu
+\varepsilon^{f}(p-eA(T))
\nonumber \\
&-&\varepsilon^{f}(p+q'-eA(T))] (\varepsilon^{f}(p-eA(T))
-\varepsilon^{f}(p+q'-eA(T))]
 \langle
c_{k}^{\dagger}c_{k+q'}f_{p+q'}^{\dagger}f_{p}\rangle (T)
\nonumber \\
&+&2U^{2}\sum_{p',q',p,q}[\varepsilon (k-eA(T))+\varepsilon
(k+q'-eA(T)) -2\mu +\varepsilon^{f}(p'-eA(T)) \nonumber \\
&-&\varepsilon^{f}(p'+q'-eA(T))] \langle
f_{p+q}^{\dagger}f_{p}f_{p'+q'}^{\dagger}f_{p'}c_{k}^{\dagger}c_{k+q'+q}\rangle
(T)
\nonumber \\
&+&2U^{2}\sum_{q',p,q,k'}[\varepsilon^{f}
(p+q-eA(T))-\varepsilon^{f}
(p+q+q'-eA(T)) \nonumber \\
&-&\varepsilon^{f}(p-eA(T)) +\varepsilon^{f}(p+q'-eA(T))] \langle
f_{p+q+q'}^{\dagger}f_{p} c_{k}^{\dagger}
c_{k+q'}c_{k'-q}^{\dagger}c_{k'}\rangle (T)
\nonumber \\
&+&2U^{2}\sum_{p',q',p,q}[\varepsilon (k+q+q'-eA(T))-\mu
-\varepsilon^{f} (p'+q'-eA(T)) +\varepsilon^{f}(p'-eA(T)) \nonumber
\\
&-&\varepsilon^{f}(p+q-eA(T))+\varepsilon^{f}(p-eA(T))] \langle
f_{p'+q'}^{\dagger}f_{p'}f_{p+q}^{\dagger}f_{p}
c_{k}^{\dagger}c_{k+q+q'}\rangle (T)
\nonumber \\
&+&2U^{3}\sum_{p',q',p,q,P,Q}\langle
f_{P+Q}^{\dagger}f_{P}f_{p'+q'}^{\dagger}f_{p'}
f_{p+q}^{\dagger}f_{p}c_{k}^{\dagger}c_{Q+k+q+q'}\rangle (T) .
\label{mukresults3}
\end{eqnarray}
The corresponding local lesser moments are
\begin{eqnarray}
\mu_{0}^{<}(T)&=&2n_{c}(T), \label{muloc<result0}\\
\mu_{1}^{<}(T)&=&-2\sum_{i,j}{\tilde t}_{ij}\langle
c_{i}^{\dagger}c_{j}\rangle -2\mu n_{c}
+2U\sum_{i}\langle f_{i}^{\dagger}f_{i}c_{i}^{\dagger} c_{i}\rangle , \label{muloc<result1}\\
\mu_{2}^{<}(T)&=& 2\sum_{i,l,j}{\tilde t}_{il}{\tilde t}_{lj}\langle
c_{i}^{\dagger}c_{j}\rangle+4\mu \sum_{i,j}{\tilde t}_{ij}\langle
c_{i}^{\dagger}c_{j}\rangle+2\mu^{2}n_{c}(T)-2U\sum_{i,j}\left[
{\tilde t}_{ij}\langle
f_{i}^{\dagger}f_{i}c_{i}^{\dagger}c_{j}\rangle +{\tilde
t}_{ji}\langle f_{i}^{\dagger}f_{i}c_{j}^{\dagger}c_{i}\rangle
\right] \nonumber \\
&~&+2U(U-2\mu )\sum_{i}\langle
f_{i}^{\dagger}f_{i}c_{i}^{\dagger}c_{i}\rangle ,
\label{muloc<result2}
\end{eqnarray}
\begin{eqnarray}
\mu_{3}^{<}(T)=&-&2\sum_{i,j,l,m}{\tilde t}_{il}{\tilde
t}_{lm}{\tilde t}_{mj}\langle c_{i}^{\dagger}c_{j}\rangle (T)
-6\mu\sum_{i,j,l}{\tilde t}_{il}{\tilde t}_{lj}\langle
c_{i}^{\dagger}c_{j}\rangle (T) -6\mu^{2}\sum_{i,j}{\tilde
t}_{ij}\langle c_{i}^{\dagger}c_{j}\rangle (T)
-2\mu^{3}\sum_{i}\langle c_{i}^{\dagger}c_{i}\rangle (T) \nonumber \\
&+&2U\sum_{i,l,j}{\tilde t}_{il}{\tilde t}_{lj}\langle
c_{i}^{\dagger}c_{j}f_{j}^{\dagger}f_{j}\rangle (T)
+2(3U\mu^{2}-3U^{2}\mu +U^{3})\sum_{i}\langle
f_{i}^{\dagger}f_{i}c_{i}^{\dagger}c_{i}\rangle (T) \nonumber \\
&+&6U\mu\sum_{i,j}{\tilde t}_{ij}\langle
c_{i}^{\dagger}c_{j}f_{j}^{\dagger}f_{j}\rangle (T) +6U\mu
\sum_{i,j}{\tilde t}_{ij}\langle
c_{i}^{\dagger}c_{j}f_{i}^{\dagger}f_{i}\rangle (T)
-2U^{2}\sum_{i,j}{\tilde t}_{ij}\langle f_{j}^{\dagger}f_{j}
c_{i}^{\dagger}c_{j}\rangle (T) \nonumber \\
&-&2U^{2}\sum_{i,j}{\tilde t}_{ij}\langle
f_{i}^{\dagger}f_{i}c_{i}^{\dagger}c_{j}\rangle (T)
+2U\sum_{i,l,j}{\tilde t}_{ij}{\tilde t}_{jl}\langle
c_{i}^{\dagger}c_{l}f_{j}^{\dagger}f_{j}\rangle (T) +2U
\sum_{i,l,j}{\tilde t}_{il}{\tilde t}_{lj}\langle
c_{i}^{\dagger}c_{j}f_{i}^{\dagger}f_{i}\rangle (T) \nonumber \\
&-&2U^{2} \sum_{i,j}{\tilde t}_{ij}\langle
f_{j}^{\dagger}f_{j}f_{i}^{\dagger}f_{i}c_{i}^{\dagger}c_{j}\rangle
(T)\nonumber \\
&+&2U\sum_{i,j,l}{\tilde t}_{ij}{\tilde t}_{il}^{f}\langle
c_{i}^{\dagger}c_{j}f_{i}^{\dagger}f_{l}\rangle (T)
-2U\sum_{i,j,l}{\tilde t}_{ij}{\tilde t}_{li}^{f}\langle
c_{i}^{\dagger}c_{j}f_{l}^{\dagger}f_{i}\rangle (T)
+2U\mu\sum_{i,j}{\tilde t}_{ij}^{f}\langle
c_{i}^{\dagger}c_{i}f_{i}^{\dagger}f_{j}\rangle (T)\nonumber \\
&-&2U\mu\sum_{i,j}{\tilde t}_{ji}^{f}\langle
c_{i}^{\dagger}c_{i}f_{j}^{\dagger}f_{i}\rangle (T)
+2U\sum_{i,j,l}{\tilde t}_{il}^{f}{\tilde t}_{lj}^{f}\langle
c_{i}^{\dagger}c_{i}f_{i}^{\dagger}f_{j}\rangle (T)
+2U\sum_{i,j,l}{\tilde t}_{il}^{f}{\tilde t}_{lj}^{f}\langle
c_{j}^{\dagger}c_{j}f_{i}^{\dagger}f_{j}\rangle (T) \nonumber \\
&-&4U\sum_{i,j,l}{\tilde t}_{ji}^{f}{\tilde t}_{il}^{f}\langle
c_{i}^{\dagger}c_{i}f_{j}^{\dagger}f_{l}\rangle (T)
-2U^{2}\sum_{i,j}{\tilde t}_{ij}^{f}[\langle
f_{i}^{\dagger}f_{i}f_{i}^{\dagger}f_{j}c_{i}^{\dagger}c_{i}\rangle
(T) -\langle
f_{j}^{\dagger}f_{j}f_{i}^{\dagger}f_{j}c_{j}^{\dagger}c_{j}\rangle
(T) ]
\nonumber \\
&-&4U^{2}\sum_{i,j}{\tilde t}_{ij}^{f}\langle
f_{i}^{\dagger}f_{j}c_{i}^{\dagger}c_{i}c_{j}^{\dagger}c_{j}\rangle
(T) +4U^{2}\sum_{i,j}{\tilde t}_{ij}^{f}\langle
f_{i}^{\dagger}f_{j}c_{j}^{\dagger}c_{j}\rangle (T),
\label{localmu<muresult3}
\end{eqnarray}
where ${\tilde t}_{ij}=t_{ij}\exp[-i\int_{R_{j}}^{R_{i}}{\bf A}({\bf
r},t)d{\bf r}]$.

Contrary to the case of the retarded moments, even the zeroth and
the first two local lesser moments in
Eqs.~(\ref{muloc<result0})-(\ref{muloc<result2}) cannot be expressed
solely in terms of the model parameters, and they depend on different
correlation functions. Therefore, in order to check the accuracy of
calculations in the lesser case, one can only compare the numerical
results for the lesser moments obtained by direct calculations by
using Eq.~(\ref{munR<}) with the corresponding numerical results
obtained by the evaluation of the Green function time derivatives in
Eq.~(\ref{munR<2}). However, the results in
Eqs.~(\ref{muloc<result0})-(\ref{localmu<muresult3}) still contain
practical importance because they provide a simple way to calculate
combinations of different correlation functions. The reason for this is due to the
fact that the correlation functions on the right hand side of
Eqs.~(\ref{muloc<result0})-(\ref{localmu<muresult3}) can be
expressed in terms of the local lesser Green functions and their
time derivatives by using Eq.~(\ref{munR<2}), the equation of motion
and/or the Dyson equations for the Green functions [see
Eqs.~(\ref{Dysonnonequilibrium}), (\ref{DysonGR})-(\ref{DysonGK})
below]. For example, as shown in Ref.~\onlinecite{spectral}, we can
connect the average potential energy with the Green functions and
self-energies:
\begin{eqnarray}
U\left\langle f_i^\dagger f_ic_i^\dagger c_i\right\rangle&=&-i
\sum_{\bf k}\left [ i\frac{\partial}{\partial
t_1}+\mu-\epsilon\left({\bf k}- \frac{e{\bf A}(t_1)}{\hbar
c}\right)\right ] G^<_{\bf k}(t_1,t_2)\Biggr |_{t_2=t_1}
\nonumber\\
&=&-i\sum_{\bf k}\int dt\left [ \Sigma^R_{\bf k}(t_1,t)G^<_{\bf
k}(t,t_1) +\Sigma^<_{\bf k}(t_1,t)G^A_{\bf k}(t,t_1)\right ] ,
\label{eq: corr_noneq}
\end{eqnarray}
a generalization of the well-known equilibrium result.

\section{Spectral moments for the retarded self-energy}
\label{Sigmamoments}

It is possible to derive expressions for the lowest retarded
self-energy moments, by using the Dyson equation, which connects the retarded Green function and self-energy, and the results for the
retarded Green function moments derived in the previous Section.

In order to derive the nonequilibrium Dyson equation for the
retarded Green function, it is convenient to write down the Dyson equation for
the contour-ordered lattice Green function in the Larkin-Ovchinnikov representation,
where all the time arguments are defined on the real branch of the
time contour:
\begin{equation}
{\hat G}_{{\bf k}}(t_{1},t_{2})={\hat G}_{{\bf
k}}^{0}(t_{1},t_{2})+\int_{-\infty}^{\infty}
dt_{3}\int_{-\infty}^{\infty} dt_{4}{\hat G}_{{\bf
k}}^{0}(t_{1},t_{3}) {\hat \Sigma}_{{\bf k}} (t_{3},t_{4}){\hat
G}_{{\bf k}}(t_{4},t_{2}) \label{Dysonnonequilibrium}
\end{equation}
and all the Green functions and self-energy functions are $2\times 2$ matrices
\begin{equation}
{\hat G}_{{\bf k}}(t_{1},t_{2})= \left(
\begin{array}{c}
G_{{\bf k}}^{R}(t_{1},t_{2})\\
0
\end{array}
\begin{array}{c}
G_{{\bf k}}^{K}(t_{1},t_{2})\\
G_{{\bf k}}^{A}(t_{1},t_{2})
\end{array}
\right) , \label{Gbasis}
\end{equation}
\begin{equation}
{\hat \Sigma}_{{\bf k}}(t_{1},t_{2})= \left(
\begin{array}{c}
\Sigma_{{\bf k}}^{R}(t_{1},t_{2})\\
0
\end{array}
\begin{array}{c}
\Sigma_{{\bf k}}^{K}(t_{1},t_{2})\\
\Sigma_{{\bf k}}^{A}(t_{1},t_{2})
\end{array}
\right) , \label{Sigmabasis}
\end{equation}
with matrix elements which consist of the retarded, advanced
\begin{equation}
G_{\bf k}^{A}(t_{1},t_{2}) =i\theta (t_{2}-t_{1})\left\langle
\left\{ c_{{\bf k}H}^{}(t_{1}),c_{{\bf
k}H}^{\dagger}(t_{2})\right\}\right\rangle , \label{GA}
\end{equation}
and the Keldysh
\begin{equation}
G_{\bf k}^{K}(t_{1},t_{2}) =-i\left\langle \left[ c_{{\bf
k}H}^{}(t_{1}),c_{{\bf k}H}^{\dagger}(t_{2}) \right] \right\rangle
\label{GK}
\end{equation}
components (and similarly for the self-energy).  The function ${\hat
G}_{{\bf k}}^{(0)}$ in Eq.~(\ref{Dysonnonequilibrium}) is the
electron Green function in the noninteracting case ($U=0$, but with
$E\ne 0$ for the nonequilibrium case). The expression for
this function can be obtained analytically (see, for example,
Refs.~\onlinecite{Jauho} and \onlinecite{Turkowski}).

The nonzero matrix components of the Dyson
equation~(\ref{Dysonnonequilibrium}) can be written in the following
form:
\begin{eqnarray}
G_{\bf k}^{R}(t_{1},t_{2})&=&G_{\bf k}^{R0}(t_{1},t_{2}) + [G_{\bf
k}^{R0}\Sigma_{{\bf k}}^{R}G_{\bf k}^{R}](t_{1},t_{2}) ,
\label{DysonGR}\\
G_{\bf k}^{A}(t_{1},t_{2})&=&G_{\bf k}^{A0}(t_{1},t_{2}) +[G_{\bf
k}^{A0}\Sigma_{{\bf k}}^{A}G_{\bf k}^{A}](t_{1},t_{2}) ,
\label{DysonGA}\\
G_{\bf k}^{K}(t_{1},t_{2})&=&[1+G_{\bf k}^{R}\Sigma_{{\bf
k}}^{R}]G_{\bf k}^{K0}[1+\Sigma_{\bf k}^{A}G_{\bf k}^{A}]
(t_{1},t_{2})+[G_{\bf k}^{R}\Sigma_{{\bf k}}^{K}G_{\bf
k}^{A}](t_{1},t_{2}) \label{DysonGK}
\end{eqnarray}
where we suppressed integrations over internal time variables implied by the
continuous matrix operator multiplications.

In order to find the retarded self-energy spectral moments, one
only needs
Eq.~(\ref{DysonGR}). It is convenient to
rewrite this equation in a combined frequency-average time
representation
\begin{eqnarray}
G_{{\bf k}}^{R}(T,\omega )&=& G_{{\bf k}}^{R0}(T,\omega )+\int
d{\bar T}\int d{\bar t} \int
d\Omega \int d\nu e^{-i\Omega {\bar t}} e^{i\nu {\bar T}} \nonumber \\
&\times&G_{{\bf k}}^{R0}\left( T+\frac{{\bar T}}{2}+\frac{{\bar
t}}{4},\omega +\Omega +\frac{\nu}{2}\right) \Sigma_{{\bf k}}^{R}
\left( T+{\bar T},\omega +2\Omega \right)G_{{\bf k}}^{R}\left(
T+\frac{{\bar T}}{2}-\frac{{\bar t}}{4},
\omega +\Omega -\frac{\nu}{2}\right), \nonumber \\
\label{Dysonnonequilibrium2}
\end{eqnarray}
where we restored the internal time/frequency integrations.

Similar to the equilibrium case\cite{Potthoff,Potthoff2,Roesch}, one
can expand the Green functions and the self-energies at large values
of the frequency $\omega$ in terms of the corresponding moments:
\begin{eqnarray}
G_{{\bf k}}^{R}(T,\omega )&=&\sum_{m=0}^{\infty}\frac{
\mu_{m}^{R}({\bf k},T)}{\omega^{m+1}}, \label{Green functionexpansionT}
\\
\Sigma_{{\bf k}}^{R}(T,\omega )&=&\Sigma_{{\bf k}}^{R}(T,\omega
=\infty )+\sum_{m=0}^{\infty}\frac{C_{m}^{R}({\bf
k},T)}{\omega^{m+1}}, \label{SigmaexpansionT}
\end{eqnarray}
where the moments $\mu_{m}^{R}({\bf k},T)$ and $C_{m}^{R}({\bf
k},T)$ correspond to the retarded Green function and self-energy in
Eq.~(\ref{DysonGR}). In particular, we have
\begin{eqnarray}
C_{m}^{R}({\bf k},T)=-\frac{1}{\pi}{\rm Im}\int_{-\infty}^{\infty}dt e^{i\omega
t}\omega^{n} \Sigma_{{\bf
k}}^{R}(T,t). \label{Sigma_moments}
\end{eqnarray}
The large-$\omega$ expansions in Eqs.~(\ref{Green functionexpansionT}) and (\ref{SigmaexpansionT})
can be obtained by using the following spectral identities (valid for retarded functions that decay rapidly enough for large relative time):
\begin{eqnarray}
G_{{\bf k}}^{R}(T,\omega )=-\frac{1}{\pi}\int_{-\infty}^{\infty}d\omega '
\frac{{\rm Im}G_{{\bf k}}^{R}(T,\omega ')}{\omega -\omega '},
\label{Green function_A}
\end{eqnarray}
\begin{eqnarray}
\Sigma_{{\bf k}}^{R}(T,\omega )=-\frac{1}{\pi}\int_{-\infty}^{\infty}d\omega '
\frac{{\rm Im}\Sigma_{{\bf k}}^{R}(T,\omega ')}{\omega -\omega '} +\Sigma_{\bf k}^R(T,\omega=\infty),
\label{Sigma_A}
\end{eqnarray}
where we take $\omega$ large enough that the Green's function and self-energy on the l.~h.~s.~are real.
In fact, by making expansions in powers of $(1/\omega)$ on the right
hand sides of Eqs.~(\ref{Green function_A}) and (\ref{Sigma_A}) and by using the
moment definitions in Eqs.~(\ref{munR<}) and (\ref{munR<2}), one can
obtain the expansions in Eqs.~(\ref{Green functionexpansionT}) and
(\ref{SigmaexpansionT}). The self-energy expansion in
Eq.~(\ref{SigmaexpansionT}) contains a frequency-independent term
$\Sigma_{{\bf k}}^{R}(T,\omega =\infty )$, which corresponds to the
mean-field term of the self-energy [see Eq.~(\ref{Ck0T}) below]; this form arises
because the self-energy generically approaches
a real constant nonzero value as $|\omega|\rightarrow\infty$.

Then, one can insert these expansions into
Eq.~(\ref{Dysonnonequilibrium2}) and consider separately the terms,
which have the same order in $(1/\omega )$. In order to do this, it
is necessary to expand all the functions under the integrals in
powers of $(1/\omega )$. For example,
\begin{eqnarray}
G_{{\bf k}}^{R}(T+\frac{{\bar T}}{2}-\frac{{\bar t}}{4}, \omega
+\Omega -\frac{\nu}{2})&=&\sum_{m=0}^{\infty}\frac{ \mu_{m}^{R}({\bf
k},T+\frac{{\bar T}}{2}-\frac{{\bar t}}{4})}{(\omega +\Omega
-\frac{\nu}{2})^{m+1}}
\nonumber \\
&=&\sum_{m=0}^{\infty}\frac{\mu_{m}^{R} ({\bf k},T+\frac{{\bar
T}}{2}-\frac{{\bar t}}{4})}{\omega^{m+1}} \frac{1}{(1+(\Omega
-\frac{\nu}{2})/\omega )^{m+1}}
\nonumber \\
&=&\sum_{m=0}^{\infty}\frac{\mu_{m}^{R} ({\bf k},T+\frac{{\bar
T}}{2}-\frac{{\bar t}}{4})}{\omega^{m+1}} \left( 1-\frac{\Omega
-\frac{\nu}{2}}{\omega}+... \right)^{m+1} . \label{1/w}
\end{eqnarray}

To calculate the frequency-independent term and the zeroth and the
first spectral moments for the retarded self-energy, it is necessary
to make an expansion of the functions in powers of $1/\omega$ in
Eq.~(\ref{Dysonnonequilibrium2}) up to fourth order. All the time
and frequency integrals in Eq.~(\ref{Dysonnonequilibrium2}) can be
easily performed, and we get the following equations which connect
the Green functions and self-energy spectral moments:
\begin{eqnarray}
\mu_{0}^{R}({\bf k},T)&=&{\tilde \mu}_{0}^{R}({\bf k},T), \label{relation0}\\
\mu_{1}^{R}({\bf k},T)&=&{\tilde \mu}_{1}^{R}({\bf k},T)+{\tilde
\mu}_{0}^{R}({\bf k},T)\Sigma_{{\bf k}}^{R}(T,\omega =\infty
)\mu_{0}^{R}({\bf k},T) ,
\label{relation1}\\
\mu_{2}^{R}({\bf k},T)&=&{\tilde \mu}_{2}^{R}({\bf k},T)+{\tilde
\mu}_{0}^{R}({\bf k},T)\Sigma_{{\bf k}}^{R}(T,\omega =\infty
)\mu_{1}^{R}({\bf k},T) +{\tilde \mu}_{0}^{R}({\bf
k},T)C_{0}^{R}({\bf k},T)\mu_{0}^{R}({\bf k},T)
\nonumber \\
&~& +{\tilde \mu}_{1}^{R}({\bf k},T)\Sigma_{{\bf k}}^{R}(T,\omega
=\infty )
\mu_{0}^{R}({\bf k},T),\label{relation2}\\
\mu_{3}^{R}({\bf k},T)&=&{\tilde \mu}_{3}^{R}({\bf k},T)+{\tilde
\mu}_{0}^{R}({\bf k},T)\Sigma_{{\bf k}}^{R}(T,\omega =\infty
)\mu_{2}^{R}({\bf k},T) +{\tilde \mu}_{0}^{R}({\bf
k},T)C_{0}^{R}({\bf k},T)\mu_{1}^{R}({\bf k},T)
\nonumber \\
&~& +{\tilde \mu}_{0}^{R}({\bf k},T)C_{1}^{R}({\bf k},T)
\mu_{0}^{R}({\bf k},T) +{\tilde \mu}_{1}^{R}({\bf k},T)\Sigma_{{\bf
k}}^{R}(T,\omega =\infty )\mu_{1}^{R}({\bf k},T)
\nonumber \\
&~& +{\tilde \mu}_{1}^{R}({\bf k},T)C_{0}^{R}({\bf k},T)
\mu_{0}^{R}({\bf k},T) +{\tilde \mu}_{2}^{R}({\bf k},T)\Sigma_{{\bf
k}}^{R}(T,\omega =\infty )\mu_{0}^{R}({\bf k},T), \label{relation3}
\end{eqnarray}
where the matrix ${\tilde \mu}_{n}^{R}({\bf k},T)$ is the $n$th
spectral moment of the retarded Green function in the {\it noninteracting case}. One
can straightforwardly derive expressions for the retarded
self-energy moments from Eqs.~(\ref{relation0})-(\ref{relation3}) by
using the results in Eqs.~(\ref{mukRresults02})-(\ref{mukRresult3}) for
the retarded Green function moments. After some long algebra, we find
\begin{eqnarray}
\Sigma_{{\bf k}}^{R}(T,\omega =\infty )&=&Un_{f} , \label{Ck0T}
\\
C_{0}^{R}({\bf k},T)&=& n_{f}(1-n_{f})U^{2} ,\label{Ck1T}
\\
C_{1}^{R}({\bf k},T)&=& U^{2}n_{f}(1-n_{f})[U(1-n_{f})-\mu ]
\nonumber \\
&+&U^{2}\sum_{p,q}[\varepsilon^{f}(p+q-eA(T))-2\varepsilon^{f}(p-eA(T))
+\varepsilon^{f}(p-q-eA(T))] \langle f_{p}^{\dagger}f_{p}\rangle (T)
\nonumber \\
&-&U^{2}\sum_{p,q,q'}[\varepsilon^{f}(p+q-eA(T))-\varepsilon^{f}(p+q+q'-eA(T))
\nonumber \\
&-&\varepsilon^{f}(p-eA(T)) +\varepsilon^{f}(p+q'-eA(T))] \langle
f_{p+q+q'}^{\dagger}f_{p}c_{k-q}^{\dagger}c_{k+q'}\rangle (T)
\nonumber \\
&+&U^{2}\sum_{p,p',q}[\varepsilon (k+q-eA(T))-\varepsilon (k-eA(T))
+\varepsilon^{f}(p'-eA(T)) -\varepsilon^{f}(p'-q-eA(T))
\nonumber \\
&+&2\varepsilon^{f} (p-eA(T))-2\varepsilon^{f}(p+q-eA(T))] \langle
f_{p'-q}^{\dagger}f_{p'}f_{p+q}^{\dagger}f_{p}\rangle (T).
\label{Ck2T}
\end{eqnarray}
The expressions for the local moments are
\begin{eqnarray}
C_{0}^{R}(T)&=&n_{f}(1-n_{f})U^{2}.\label{C1Tloc}
\\
C_{1}^{R}(T)&=&U^{2}n_{f}(1-n_{f})[U(1-n_{f})-\mu ] \nonumber
\\
&~&+2U^{2}\sum_{ij}{\tilde t}_{ij}^{f}\langle
f_{i}^{\dagger}f_{j}\rangle -U^{2}\sum_{ij}{\tilde
t}_{ij}^{f}\left(\langle
f_{i}^{\dagger}f_{j}c_{j}^{\dagger}c_{j}\rangle +\langle
f_{i}^{\dagger}f_{j}c_{i}^{\dagger}c_{i}\rangle \right) \nonumber \\
&~&+U^{2}\sum_{ij}{\tilde t}_{ij}^{f}\left( -\langle
f_{i}^{\dagger}f_{j}f_{i}^{\dagger}f_{i}\rangle +\langle
f_{i}^{\dagger}f_{j}f_{j}^{\dagger}f_{j}\rangle -2\langle
f_{i}^{\dagger}f_{i}f_{i}^{\dagger}f_{j}\rangle +2\langle
f_{j}^{\dagger}f_{j}f_{i}^{\dagger}f_{j}\rangle\right) .
\label{C2Tloc}
\end{eqnarray}
It is worthwhile to notice that the local retarded self-energy
moments are time independent (except for the first moment in the
case of the Hubbard model, for which we are not sure about the time
dependence). This may be a surprising result for the Hubbard model, since the second-order perturbation theory is frequency-dependent, but the total weight of the self-energy reamins constant and depends just on the electron densities and the interaction. Other interesting observations are that the mean-field
term $\Sigma_{{\bf k}}^{R}(T,\omega =\infty )$ is equal to the first
order (Hartree-Fock) term of the self-energy in the expansion in
$U$, and that the zeroth moment
 corresponds to the zeroth moment of the imaginary part of
$\Sigma^{R}(\omega, T)$ in the truncated second-order perturbation
expansion\cite{PT} (for the Falicov-Kimball model case). This is in agreement with a result of
Ref.~\onlinecite{Deisz}, where it was shown that in equilibrium the
exact coefficient of the term proportional to $1/\omega_{n}$ in the
large Matsubara frequency expansion of the electron self-energy of
the Hubbard model can be obtained from the second-order skeleton
diagram for the exact Green function. Finally, it was shown in
Ref.~\onlinecite{Freericks_sigma_delta}, in the insulating phase,
that the imaginary part of the $d\rightarrow\infty$ equilibrium
retarded self-energy acquires an additional term proportional to
$\delta(\omega)$ in the frequency representation (at half filling; away from half filling a delta function appears but not at $\omega=0$). In particular, in
the case of the Falicov-Kimball model, the weight of the delta
function term is equal to $-\pi [U^{2}n_{f}(1-n_{f})-1/2]$ and it
produces a term that requires special care to include in the zeroth self-energy moment, when one performs the integration over frequency of the self-energy. Note that the delta function implies that the finite-frequency integration of the zeroth self-energy moment remains fixed at 0.5 for the Falicov-Kimball model in the insulating phase at half-filling, and all of the additional spectral weight comes from the delta function piece at $\omega=0$. Away from half filling the delta function typically contributes to all moments because it appears at a finite frequency. In the
nonequilibrium case, the situation is more complicated, because we
cannot prove that such a term is also present in this case. To see whether such a term is present, one needs to examine the large relative-time limit of the nonequilibrium retarded self-energy, which would have a constant term equal to the weight of the delta function when the delta function appears at $\omega=0$ (and would be a term proportional to $\exp i\omega t$ when the delta function is at a finite frequency), but we do find good overall agreement for the sum rules, so this issue is not important in verifying the accuracy (when one performs calculations in the time representation).

Unfortunately, it is impossible to derive analogous expressions for
the lesser self-energy spectral moments
\begin{eqnarray}
C_{m}^{<}({\bf k},T)=\frac{1}{\pi}\int_{-\infty}^{\infty}d\omega \omega^{n}
 {\rm Im}\Sigma_{{\bf k}}^{<}(T,\omega ),
\label{Sigma<_moments}
\end{eqnarray}
since in this case the expansions similar to
Eqs.~(\ref{Green functionexpansionT}) and (\ref{SigmaexpansionT}) do not exist. In
fact, the representations in Eqs.~(\ref{Green function_A}) and (\ref{Sigma_A}), which
lead to Eqs.~(\ref{Green functionexpansionT}) and (\ref{SigmaexpansionT}), are not
valid in the cases of the lesser Green function and self-energy, because the lesser
functions are pure imaginary and hence not analytic. Note that we could try to define
an auxiliary Green's function that has the imaginary part of the lesser Green's function
and a real part determined by the spectral function defined by the integral of the imaginary part,
but doing so does not produce any new results for the spectral moments of the lesser self-energy.

An alternate approach is to express the lesser self-energy
in terms of the retarded Green function and self-energy by using the system of
Dyson equations in Eqs.~(\ref{DysonGR})-(\ref{DysonGK}) and the equation
which connects the lesser Green function with the retarded, advanced and Keldysh
Green functions,
\begin{equation}
G_{\bf k}^{<}(t_{1},t_{2}) = \frac{1}{2}\left( G_{\bf
k}^{K}(t_{1},t_{2}) -G_{\bf k}^{R}(t_{1},t_{2}) +G_{\bf
k}^{A}(t_{1},t_{2}) \right) ,  \label{GlGK}
\end{equation}
and then try to express the lesser self-energy moments in terms of
moments for the retarded and lesser Green function and the retarded self-energy.
In this case, one can find the following Dyson equation for the
lesser self-energy:
\begin{eqnarray}
G_{\bf k}^{<}(t_{1},t_{2})= [1+G_{\bf k}^{R}\Sigma_{{\bf
k}}^{R}]G_{\bf k}^{<0}[1+\Sigma_{\bf k}^{A}G_{\bf k}^{A}]
(t_{1},t_{2})+[G_{\bf k}^{R}\Sigma_{{\bf k}}^{<}G_{\bf
k}^{A}](t_{1},t_{2}). \label{DysonG<}
\end{eqnarray}
Using the equations of motion for the Green functions,
\begin{eqnarray}
\left[ \delta (t-t_{1})\left(i\frac{\partial}{\partial t_{1}}
-\varepsilon (k-A(t_{1}))+\mu \right) -\Sigma_{\bf k}^{R}(t,t_{1})
\right]
G_{\bf k}^{R}(t_{1},t_{2})&=&\delta (t-t_{2}), \label{EqR} \\
G_{\bf k}^{A}(t_{1},t_{2})\left[ \left(i\frac{\partial}{\partial
t_{2}} +\varepsilon (k-A(t_{2}))-\mu \right)\delta (t_{2}-t')
+\Sigma_{\bf
k}^{A}(t_{2},t') \right] &=&-\delta (t_{1}-t'), \label{EqA} \\
\left[ \delta (t-t_{1})\left(i\frac{\partial}{\partial t_{1}}
-\varepsilon (k-A(t_{1}))-\mu \right) -\Sigma_{\bf k}^{<}(t,t_{1})
\right] G_{\bf k}^{<}(t_{1},t_{2})&=&0, \label{Eq}
\end{eqnarray}
one can get the following formal expression for the lesser
self-energy:
\begin{eqnarray}
\Sigma_{\bf k}^{<}(t_{1},t_{2}) &=& -\left[\delta
(t_{1}-t)\left(i\frac{\partial}{\partial t} -\varepsilon
(k-A(t))+\mu \right) -\Sigma_{\bf k}^{R}(t_{1},t) \right]G_{\bf
k}^{<}(t,t')
\nonumber \\
&~&\times\left[ \left( i\frac{\partial}{\partial t'} +\varepsilon
(k-A(t'))-\mu \right)\delta (t'-t_{2}) +\Sigma_{\bf
k}^{A}(t',t_{2})\right] . \label{Eq2}
\end{eqnarray}
Using this result, one can calculate the lesser self-energy moments
similar to what was done for the Green functions:
\begin{equation}
C_{n}^{<}(k,T)= \frac{1}{\pi}{\rm Im}\left[
\frac{1}{(-i)^{n}}\frac{\partial^{n}}{\partial t^{n}} \Sigma_{\bf
k}^{<}(T,t )\right]_{t=0^{+}}, \label{Cn<}
\end{equation}
where $T$ and $t$ are the average and the relative time coordinates.

Unfortunately, this approach also does not provide any useful results for the self-energy moments.
In fact, even in the equilibrium case,
one finds from Eqs.~(\ref{Eq2}) and (\ref{Cn<}) the following trivial
result:
\begin{eqnarray}
C_{n}^{<}(k,T)&=&\frac{1}{\pi}\int d\omega \omega^{n}f(\omega ){\rm
Im}\Sigma_{\bf k}^{R}(\omega).\label{Sigma_eq3}
\end{eqnarray}
[In order to obtain this expression, one needs to use the following
equilibrium relations: $\Sigma_{\bf k}^{A}(\omega)=\Sigma_{\bf
k}^{R*}(\omega)$ and $G_{\bf k}^{<}(\omega)=if(\omega){\rm Im}G_{\bf
k}^{R}(\omega)$]. The result in Eq.~(\ref{Sigma_eq3}) can also be
obtained directly from the equilibrium relation $\Sigma_{\bf
k}^{<}(\omega)=if(\omega ){\rm Im}\Sigma_{\bf k}^{R}(\omega)$.
Unfortunately, it is impossible to get analytical results for the
lesser self-energy moments from Eq.~(\ref{Sigma_eq3}), except in the
high-temperature limit, when they can be expressed in terms of the
retarded self-energy moments [via a series expansion for
$f(\omega)$].

Since the exact analytical results for the lesser moments cannot be
found even in the equilibrium case, one can try to make some
approximations in order to obtain them. The standard approximation
for the lesser Green function is the generalized Kadanoff-Baym (GKB)
approximation\cite{GKB}:
\begin{eqnarray}
G_{\bf k}^{<}(t_{1},t_{2})=-i\left[ G^{R}(t_{1},t_{2})G_{\bf
k}^{<}(t_{2},t_{2})-G_{\bf k}^{<}(t_{1},t_{1})G_{\bf
k}^{A}(t_{1},t_{2}) \right] . \label{GKB}
\end{eqnarray}
Substitution of this result into Eq.~(\ref{Eq2}) and using the equations
of motion in Eqs.~(\ref{EqR})-(\ref{Eq}) gives the following approximate
result for the lesser self-energy:
\begin{eqnarray}
\Sigma_{\bf k}^{<}(t_{1},t_{2})=-i\left[
\Sigma^{R}(t_{1},t_{2})G_{\bf k}^{<}(t_{2},t_{2})-G_{\bf
k}^{<}(t_{1},t_{1})\Sigma_{\bf k}^{A}(t_{1},t_{2}) \right] -2i\delta
(t_{1}-t_{2}) \frac{\partial n_{\bf k}(t_{2})}{\partial t_{2}},
\label{GKBSigma}
\end{eqnarray}
or in the frequency-average time representation:
\begin{eqnarray}
\Sigma_{\bf k}^{<}(T,\omega )=2i{\rm Im} \Sigma_{\bf k}^{R}(T, \omega )
n_{\bf k}(T) -2i\frac{\partial n_{\bf k}(T)}{\partial T}.
\label{GKBSigma2}
\end{eqnarray}
After summation over momentum the last term disappears, due to
conservation of the total particle number, therefore in this case
\begin{eqnarray}
C_{n}^{<}(T)=2\sum_{\bf k} C_{n}^{R}({\bf k},T)n_{\bf k}(T).
\label{GKBSigma3}
\end{eqnarray}
Since the zeroth and the first retarded self-energy moments are
momentum-independent, one can obtain the following GKB result for the
corresponding lesser moments
\begin{eqnarray}
C_{n}^{<}(T)=2C_{n}^{R}(T)n_{c}. \label{GKBSigma4}
\end{eqnarray}
The GKB approximation gives good results for the Green's function moments in the case of weakly
interacting systems. Therefore, the relation Eq.~(\ref{GKBSigma4})
should be approximately valid in this case.  There is one subtle issue with regards
to the GKB and DMFT.  In DMFT, the self-energy is local, and hence momentum independent.
But the GKB approximation to the self-energy in Eq.~(\ref{GKBSigma2}) appears to be momentum dependent.
Hence, it is not clear how accurate the local self-energy moments will be within this approximation,
but because the GKB corresponds to a mean-field-like decoupling of correlation functions
for the Green function moments\cite{spectral}, it is possible that the approximation
remains reasonable for the local self-energy, at least for weak coupling.

Thus, generally speaking, similar to lesser Green function moment case, one
cannot obtain analytical expressions for the lesser self-energy
moments. Moreover, it is even impossible to express these
moments in terms of correlation functions. Hence, in order to check
the accuracy of the numerical calculations, one can only compare the
numerical results for the moments with the numerical evaluation of
the self-energy time derivatives in Eq.~(\ref{Cn<}), which is not a stringent test.

\section{Numerical results for the Falicov-Kimball model in infinite dimensions}
\label{FalicovKimball}

In this Section, we shall use results for the local moments obtained
in Sections~\ref{Green functionmoments}-\ref{Sigmamoments} to check
the accuracy of the equilibrium and nonequilibrium numerical
solutions of the Falicov-Kimball model in the limit of
infinite dimensions. In this limit, the electron self-energy is
local\cite{metzner_1991}, which allows one to solve the
problem numerically in both
equilibrium\cite{Freericks} and nonequilibrium
cases\cite{Nashville,spectral,denver,PRL,review,freericks_current_long}. The case of infinite
dimensions is important, since many physical properties of the model
are qualitatively similar as in the 2D and 3D cases
(see, for example, Ref.~\onlinecite{Freericks}).

\begin{figure}[h]
\centering{
\includegraphics[width=1.5in,angle=270]{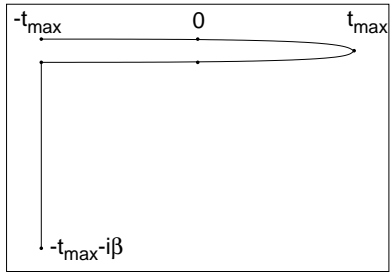}}
\caption{The complex Kadanoff-Baym-Keldysh contour for the two-time Green's
functions in the nonequilibrium case.} \label{fig: keldysh}
\end{figure}

In order to study the time-dependent properties of the model in
infinite dimensions, one needs to solve a generalized system of
nonequilibrium DMFT equations for the contour ordered Green's
function $G (t_{1},t_{2})$, self-energy
 $\Sigma (t_{1},t_{2})$ and an effective dynamical mean-field
$\lambda (t_{1},t_{2})$:
\begin{eqnarray}
G(t_{1},t_{2})&=&\sum_{{\bf k}}[G_{{\bf k}}^{(0)-1}-
\Sigma]^{-1}(t_{1}, t_{2}) ,
 \label{DMFT1}
\\
G_0(t_1,t_2)&=&[G^{-1}+\Sigma]^{-1}(t_1,t_2), \label{g0_def}
 \\
\lambda (t_{1},t_{2})
&=&G^{-1}_{0imp}(t_{1},t_{2};\mu)-G_0^{-1}(t_{1},t_{2}),
 \label{DMFT2}
\\
 G(t_{1},t_{2})
&=&(1-w_{1} )G_{0}(t_{1}, t_{2};\mu) +
w_{1}[G_{0imp}^{-1}(\mu-U)-\lambda]^{-1}(t_{1}, t_{2}),
 \label{DMFT3}
\end{eqnarray}
where all time arguments are defined on the complex Kadanoff-Baym-Keldysh
time contour (see Fig.~\ref{fig: keldysh}).
On this contour, the time increases from the top left point
($-t_{max}$) along the contour to the bottom point of the imaginary
axis ($-t_{max}-i\beta$). In Eqs.~(\ref{DMFT1})-(\ref{DMFT3}),
$G_{{\bf k}}^{(0)}(t_{1}, t_{2})$ is the noninteracting electron
Green's function in the presence of an external field and
$G_{0imp}(t_{1}, t_{2};\mu)$ is the free impurity Green function;
$\mu$ is a chemical potential and $w_{1}$ is the average
number of the $f$-electrons per site (for details, see
Refs.~\onlinecite{review} and \onlinecite{freericks_current_long}).

As mentioned in Section~\ref{Formalism}, we shall consider
the  case of a spatially uniform electric
field directed along the elementary cell diagonal, as in Eq.~({\ref{A}}). We
also assume that the system starts in equilibrium with an inverse temperature $\beta$ and then a constant electric field is turned on at time
$t=0$.

In the case of an  external field, as given in Eq.~({\ref{A}}), the free electron
spectrum [in Eq.~(\ref{freespectra})] has a simple momentum dependence:
\begin{equation}
\epsilon \left({\bf k}-\frac{e{\bf A}(t)}{\hbar c}\right)
=\cos\left(\frac{eaA(t)}{\hbar c}\right)\epsilon ({\bf k})
+\sin\left(\frac{eaA(t)}{\hbar c} \right){\bar \varepsilon} ({\bf
k}), \label{energy}
\end{equation}
where
\begin{equation}
\epsilon ({\bf k})=-2t\sum_{l}\cos (ak^{l}) \label{eps}
\end{equation}
and
\begin{equation}
{\bar \varepsilon} ({\bf k})=-2t\sum_{l}\sin (ak^{l}).
\label{bareps}
\end{equation}
are two energy functions. It is possible to show that in the case of
an infinite dimensional hypercubic lattice, the joint density of
states for these two energy functions has the following form
\cite{Schmidt}:
\begin{equation}
\rho_{2}(\epsilon , {\bar \varepsilon}) =\frac{1}{\pi
t^{*2}a^{d}}\exp\left[ -\frac{\epsilon^{2}}{t^{*2}}-\frac{{\bar
\varepsilon}^{2}}{t^{*2}} \right] , \label{rho2}
\end{equation}
where $t^{*}$ is a scaled hopping parameter, connected with the
hopping $t$ in the Hamiltonian Eq.~(\ref{H}) as $t=t^*/2\sqrt{d}$.
The momentum summation in Eq.~(\ref{DMFT1}) can be performed by
using the joint density of states Eq.~(\ref{rho2}): $\sum_{{\bf
k}}F_{{\bf k}}=\int d\epsilon \int d{\bar \varepsilon}
\rho_{2}(\epsilon , {\bar \varepsilon})F_{\epsilon, {\bar
\varepsilon}}$, since in our case the noninteracting Green's function on the
r~h~s.~of Eq.~(\ref{DMFT1}) has simple momentum-dependence, which
can be expressed in terms of the two energy functions in Eqs.~(\ref{eps})
and (\ref{bareps}). The energy integration can be performed by using
Gaussian integration\cite{Nashville,denver}. We typically use about 100 points per dimension.

In addition, one needs to choose the
proper discretization of the time contour Fig.~\ref{fig: keldysh}.
The results depend strongly on
the discretization step when the step size is not small enough. Choosing a given discretization and a $t_{max}$ determines the matrix size for the given calculations.  We typically work with general complex matrixes of size $900\times 900$ up to $5700\times 5700$.

\subsection{Equilibrium case}
\label{FalicovKimball_equilibrium}

First, we consider the equilibrium case, when there is no external
field. In this case, the system of equations
(\ref{DMFT1})-(\ref{DMFT3}) reduces to the equilibrium DMFT
equations\cite{jarrell_1992} with no average time dependence, so functions of two time
arguments can be replaced by corresponding functions of one
frequency, $F(t_{1},t_{2})\rightarrow F(\omega )$. The numerics are under good control and one can obtain quite accurate solutions. The most important numerical checks that can be performed arise from a comparison of the spectral moments calculated directly by integrating the real-frequency solutions, with results for the moments that can be determined exactly via parameters of the model for the retarded moments or by an evaluation of the relevant correlation functions using a Matsubara frequency formalism for the lesser moments.

Now we show how to calculate the required
correlation functions in
Eqs.~(\ref{muloc<result0})-(\ref{localmu<muresult3}) using
the Matsubara Green's functions.  One starts from the imaginary time-ordered Green's functions
\begin{eqnarray}
G_{ij}(\tau)=-\langle {\rm T}_{\tau} c_{i}(\tau ) c_{j}^{\dagger}(0)
\rangle , \label{Matsubara}
\end{eqnarray}
where the imaginary time-dependent operators satisfy $c_{i}(\tau
)=e^{H\tau}c_{i}(0)e^{-H\tau}$ according to the Heisenberg
representation. Because these functions are antiperiodic on the interval $[0,\beta]$, we employ a Fourier expansion in terms of the Matsubara frequencies:
\begin{eqnarray}
G_{ij}(\tau)=T\sum_{n}e^{-i\omega_{n}\tau}\int d{\bf k}e^{-i{\bf
k}({\bf R}_{i}-{\bf R}_{j})}G_{{\bf k}}(i\omega_{n}),
\label{Matsubara2}
\end{eqnarray}
where $\omega_{n}=\pi T(2n+1)$ is the fermion Matsubara frequency.
Here, the momentum-dependent Matsubara Green's function satisfies
\begin{eqnarray}
G_{{\bf k}}(i\omega_{n})=\frac{1}{i\omega_{n}+\mu -\varepsilon ({\bf
k})-\Sigma_{{\bf k}}(i\omega_{n})}, \label{Matsubara3}
\end{eqnarray}
and in DMFT the self-energy has no momentum dependence.

We start by deriving the equation of motion for the Green function in Eq.~({\ref{Matsubara})
and extracting the expression for the local four-operator correlation
function by evaluating the Green function at $\tau=0$ and removing the single-particle terms:
\begin{eqnarray}
\langle f_{i}^{\dagger}f_{i}c_{i}^{\dagger}c_{i}\rangle
=\frac{T}{U}\sum_{n,{\bf k}}\Sigma_{{\bf k}}(i\omega_{n})G_{{\bf
k}}(i\omega_{n}). \label{corr1}
\end{eqnarray}
The correlation functions for operators on
different sites, like $\langle
f_{i}^{\dagger}f_{i}c_{i}^{\dagger}c_{j}\rangle$, can be found by introducing an extra term
$-\sum_{i}h_{i}f_{i}^{\dagger}f_{i}$ with a local field $h_{i}$ into
the equilibrium Hamiltonian and then evaluating derivatives with respect to $h_i$ and taking the limit $h_i\rightarrow 0$. For example, straightforward algebra shows that
\begin{eqnarray}
\langle f_{i}^{\dagger}f_{i}c_{i}^{\dagger}c_{j}\rangle =\left[
T\frac{\partial}{\partial h_{i}}+\langle w_{i}\rangle \right]
G_{ij}(\tau =0^{-}), \label{corr2}
\end{eqnarray}
where $\langle w_{i}\rangle =\langle f_{i}^{\dagger}f_{i}\rangle
=n_{fi}$ (see Refs.~\onlinecite{spectral}, \onlinecite{Freericks2},
and \onlinecite{Freericks3} and the Appendix for details). Using these identities allows us to find explicit expressions for all of the relevant correlation functions using Green functions and self-energies determined at the Matsubara frequencies. We present
the final results for the case of the Falicov-Kimball model in
infinite dimensions, where the self-energy is momentum-independent:
\begin{eqnarray}
n_{c}&=&T\sum_{n,{\bf k}}G_{{\bf k}}(i\omega_{n}), \label{corr_k_1}\\
\sum_{i,j}t_{ij}\langle c_{i}^{\dagger}c_{j}\rangle
&=&-T\sum_{n,{\bf k}}\varepsilon ({\bf k})G_{{\bf k}}(i\omega_{n}),
\label{corr_k_2}
\\
\sum_{i,l,j}t_{il}t_{lj}\langle c_{i}^{\dagger}c_{j}\rangle
&=&T\sum_{n,{\bf k}}\varepsilon^{2} ({\bf k})G_{{\bf
k}}(i\omega_{n}), \label{corr_k_3} \\
\sum_{i,l,m,j}t_{il}t_{lm}t_{mj}\langle c_{i}^{\dagger}c_{j}\rangle
&=&-T\sum_{n,{\bf k}}\varepsilon^{3} ({\bf k})G_{{\bf
k}}(i\omega_{n}), \label{corr_k_4} \\
\sum_{i}\langle f_{i}^{\dagger}f_{i}c_{i}^{\dagger}c_{i}\rangle
&=&\frac{T}{U}\sum_{n,{\bf k}}\Sigma (i\omega_{n}) G_{{\bf
k}}(i\omega_{n}), \label{corr_k_5} \\
\sum_{i,j}t_{ij}\langle
f_{i}^{\dagger}f_{i}c_{i}^{\dagger}c_{j}\rangle
&=&\left[\sum_{i,j}t_{ji}\langle
f_{i}^{\dagger}f_{i}c_{j}^{\dagger}c_{i}\rangle\right]^{*}=
-\frac{T}{U}\sum_{n,{\bf k}}\Sigma (i\omega_{n}) \varepsilon ({\bf
k}) G_{{\bf k}}(i\omega_{n}), \label{corr_k_6} \\
\sum_{i,l,j}t_{il}t_{lj}\langle
c_{i}^{\dagger}c_{j}f_{i}^{\dagger}f_{i}\rangle
&=&\sum_{i,l,j}t_{il}t_{lj}\langle
c_{i}^{\dagger}c_{j}f_{j}^{\dagger}f_{j}\rangle =
\left[\sum_{i,l,j}t_{ij}t_{jl}\langle
c_{i}^{\dagger}c_{l}f_{j}^{\dagger}f_{j}\rangle\right]^{*}=
\frac{T}{U}\sum_{n,{\bf k}}\Sigma (i\omega_{n}) \varepsilon^{2}
({\bf k}) G_{{\bf k}}(i\omega_{n}), \nonumber \\
\label{corr_k_7}\\
\sum_{i,j}t_{ij}\langle
f_{j}^{\dagger}f_{j}f_{i}^{\dagger}f_{i}c_{i}^{\dagger}c_{j}\rangle
&=&-\frac{T}{U^{2}}\sum_{n,{\bf k}}\Sigma^{2}(i\omega_{n})
\varepsilon ({\bf k}) G_{{\bf k}}(i\omega_{n}). \label{corr_k_8}
\end{eqnarray}
We next perform the momentum summation in
Eqs.~(\ref{corr_k_1})-(\ref{corr_k_8}) to express the results in
terms of local quantities:
\begin{eqnarray}
n_{c}&=& T\sum_{n}G_{n}, \label{corr_int_1}\\
\sum_{i,j}t_{ij}\langle c_{i}^{\dagger}c_{j}\rangle
&=&T\sum_{n}\left[ 1-(i\omega_{n}+\mu-\Sigma_{n})G_{n}\right] , \label{corr_int_2}\\
\sum_{i,l,j}t_{il}t_{lj} \langle c_{i}^{\dagger}c_{j}\rangle
&=&-T\sum_{n}(i\omega_{n}+\mu-\Sigma_{n})\left[
1-(i\omega_{n}+\mu-\Sigma_{n})G_{n}\right] ,
\label{corr_int_3}\\
\sum_{i,l,m,j}t_{il}t_{lm}t_{mj}\langle c_{i}^{\dagger}c_{j}\rangle
&=&T\sum_{n}\left[\frac{1}{2}+(i\omega_{n}+\mu-\Sigma_{n})^{2}(1-(i\omega_{n}+\mu-\Sigma_{n})G_{n})\right]
, \label{corr_k_4b} \\
\sum_{i}\langle f_{i}^{\dagger}f_{i}c_{i}^{\dagger}c_{i}\rangle
&=&\frac{T}{U}\sum_{n}\Sigma_{n}G_{n}, \label{corr_int_5} \\
\sum_{i,j}t_{ij}\langle
f_{i}^{\dagger}f_{i}c_{i}^{\dagger}c_{j}\rangle
&=&\left[\sum_{i,j}t_{ji}\langle
f_{i}^{\dagger}f_{i}c_{j}^{\dagger}c_{i}\rangle\right]^{*}=
\frac{T}{U}\sum_{n}\Sigma_{n}\left[
1-(i\omega_{n}+\mu-\Sigma_{n})G_{n}\right] ,
\label{corr_int_6} \\
\sum_{i,l,j}t_{il}t_{lj}\langle
c_{i}^{\dagger}c_{j}f_{j}^{\dagger}f_{j}\rangle
&=&\left[\sum_{i,l,j}t_{il}t_{lj}\langle
c_{i}^{\dagger}c_{j}f_{i}^{\dagger}f_{i}\rangle\right]^{*}=
\sum_{i,l,j}t_{ij}t_{jl}\langle
c_{i}^{\dagger}c_{l}f_{j}^{\dagger}f_{j}\rangle \nonumber \\
&=&-\frac{T}{U}\sum_{n}\Sigma_{n}(i\omega_{n}+\mu-\Sigma_{n})\left[
1-(i\omega_{n}+\mu-\Sigma_{n})G_{n}\right] ,
\label{corr_int_7}\\
\sum_{i,j}t_{ij}\langle
f_{j}^{\dagger}f_{j}f_{i}^{\dagger}f_{i}c_{i}^{\dagger}c_{j}\rangle
&=&\frac{T}{U^{2}}\sum_{n}\Sigma^{2}(i\omega_{n})\left[
1-(i\omega_{n}+\mu-\Sigma_{n})G_{n}\right] , \label{corr_int_8}
\end{eqnarray}
where $G_{n}\equiv \sum_{{\bf k}}G_{{\bf k}}(i\omega_{n})$ and
$\Sigma_{n}\equiv \Sigma (i\omega_{n})$. These expressions can then be employed to efficiently determine the lesser moments from an independent Matsubara frequency calculation.

We find, for all cases that we consider, all of the different Green's function and self-energy moment sum rules are satisfied to essentially as high an accuracy as we want (the delta function contributions to the self-energy moments must be included to get the correct answer; this becomes complicated for particle-hole asymmetric cases when $U$ is large enough for the self-energy to have developed a pole because one needs to accurately determine the location and weight of the pole to obtain the correct sum rules).  In some cases, we need to use many Matsubara frequencies in the summations to achieve sufficient accuracy, or we need to have a small frequency grid spacing for the real-frequency Green's functions. The sum rules hold in the case of half-filling and away from particle-hole symmetry and they hold equally well for metallic and insulating cases.

\subsection{Nonequilibrium case}
\label{FalicovKimball_nonequilibrium}

In this Subsection, we compare the numerical results for the moments (at half filling)
with exact analytical results obtained in the case when a constant
electric field is turned on at time $t=0$. Since we calculate the contour-ordered self-energy, we need to extract the correct retarded quantities to compare with the moments that do not depend on correlation functions (which we have no independent way to evaluate). This is simple to do for the Green's functions.  For the self-energies care is needed.  The constant term in the self-energy in the frequency representation becomes an equal time delta function in the time formalism. The zeroth moment corresponds to the equal-time retarded self-energy (most easily found by taking the difference of the greater and lesser self-energies) and the first moment is found from the first derivative.  We need to evaluate the derivative carefully, because we need to remove the delta-function piece first.  We handle this instead by using linear extrapolation from finite relative times to the vanishing relative-time limit, so we do not need the data at equal times to find the derivative. More sophisticated techniques would be needed to find the higher moments, but we don't need those here.

In general, the self-energy moments are satisfied to very high accuracy, even if the step size is large.  Errors are often less than 0.1\%, which is much lower than what one finds for the Green's function moments (where we often need to work hard to get errors below the 1\% level\cite{denver,PRL,freericks_current_long}). We can extrapolate the results to the limit $\Delta t\rightarrow 0$, which produces even higher accuracy.  The results are most accurate for the constant piece to the self-energy.  Then the zeroth moment, and finally the first moment.  But the results of our investigations indicate that the Green's function moments are a much more accurate test of the accuracy of the solutions than the self-energy moments.  While we could show similar scaling plots as were created for the Green's function moments\cite{denver}, it does not seem to be necessary because the improved accuracy is so much better for the self-energies that one does not learn too much from such an exercise.

\section{Conclusions}
\label{Conclusions}

In this work we have shown how to extend the Green's function moment sum rules to third order for both the Hubbard and the Falicov-Kimball models and used these moments to examine the retarded self-energy moments through first order. Our analysis holds both for equilibrium and nonequilibrium situations.   We find for the Falicov-Kimball model that the moment sum rules remain time independent in nonequilibrium, which is a surprising result.  In the case of the Hubbard model, it appears that the third order moments will be time dependent, but we cannot explicitly confirm this.  When we compare the sum rules to numerical calculations for the Falicov-Kimball model with DMFT, we find excellent agreement both in equilibrium and in nonequilibrium.   In fact, the Green's function sum rules are a much better indicator of overall accuracy than the self-energy sum rules.

The sum rules are only relevant for quantitative comparisons of retarded functions.  In the case of lesser functions, we are able to make comparisons of the Green's function sum rules to the relevant correlation functions evaluated with a Matsubara frequency formalism when the system is in equilibrium, but we cannot extend that approach to the nonequilibrium case.  We are unable, even in equilibrium, to find any useful sum rules for the lesser self-energy.  Instead we find just trivial relationships that arise from the definitions of these quantities (which are well known in equilibrium and unknown in nonequilibrium).

In the future, we will examine how these sum rules can be extended to inhomogeneous situations, with relevance to inhomogeneous DMFT (and other techniques) as applied to mutlilayered nanostructures or ultracold atomic systems in a harmonic trap. In addition, utilizing these sum rules can allow one to obtain more accurate results for the high-frequency limit of the Green's functions, self-energies and dynamical mean fields. We will illustrate this use in another publication, which allows one to employ a minimal number of Matsubara frequencies yet maintain high accuracy of solutions.

\section*{Acknowledgments}

V.T. would like to acknowledge support by the National Science
Foundation under grants numbered DMR-0553485 and DMR-0705266.
J.~K.~F. acknowledges support from the N.~S.~F.~under grant number DMR-0705266.  Supercomputer time was provided by CAP phase II projects at the ERDC and ARSC supercomputer centers of the HPCMP and by NASA under a National Leadership in Computing Systems grant.

\appendix

\section{Calculation of equilibrium correlation functions using the Matsubara frequency formalism}

In this Appendix, we present details of the derivation of the
correlation functions in
Eqs.~(\ref{corr_k_5})-(\ref{corr_k_8}). The expression in Eq.~(\ref{corr1})
for the first correlation function in Eq.~(\ref{corr_k_5}) can also be determined by introducing a fictitious field $-\sum_i h_i f^\dagger_if^{}_i$ into the Hamiltonian and taking derivatives with respect to $h_i$ and then setting all $h_i=0$. Each derivative with respect to an $h_i$ brings down an operator $f^\dagger_if^{}_i$ into the operator average (plus a correction term when the derivative acts on the partition function). This approach is more general than the equation of motion approach used to derive Eq.~(\ref{corr1}) and will allow us
 to derive expressions for the other
correlation functions in Eqs.~(\ref{corr_k_6})-(\ref{corr_k_8}). As
shown in Ref.~\onlinecite{Freericks3}, a correlation function
that contains a product of two $c$-electron operators and one
$f$-electron number operator can be expressed in terms of a
derivative of the $c$-electron Green function with respect to the fictitious field [for example, Eq.~(\ref{corr2})].

In order to explicitly calculate the fictitious field derivative of the
Green function, one uses the standard trick of writing $G=GG^{-1}G$ so that derivatives of $G$ are replaced by derivatives of $G^{-1}$ which involves a derivative of the self-energy  (see
Ref.~\onlinecite{Freericks3}). Because we have added the fictitious fields to the Hamiltonian, and they are not translationally invariant, we lose translational invariance in the system prior to taking the derivatives (it is restored once we set $h_i=0$). Hence, we need to work in real space rather than momentum space, and we need to allow the dynamical mean fields and the self-energies to have a site dependence.  This implies that we can write the local Green function at site $i$ via
\begin{equation}
G_{ii}(i\omega_{n})=\frac{1}{i\omega_{n}+\mu-\lambda_{i}(i\omega_{n})-\Sigma_{i}(i\omega_{n})}
\label{DMFT}
\end{equation}
in the Matsubara frequency representation.

Now consider the case where we add an $h$-field only at site $i$. Since the $h$ field will modify $n_{fi}$, the Green function and self-energy at site $i$ are changed by $h_i$. What about the Green function and self-energy on neighboring sites? Using the Dyson equation, one can show that the change in the Green function at site $j$, $\delta G_{jj}(i\omega_n)$, is equal to
\begin{equation}
 \delta G_{jj}(i\omega_n)=G_{ji}(i\omega_n)|_{h_i=0}\delta \Sigma_i(i\omega_n)G_{ij}(i\omega_n)|_{h_i=0}.
\label{eq: inhomog_dyson}
\end{equation}
But $G_{ij}$ is proportional to the hopping $t$ raised to the power equal to smallest number of hops between site $i$ and site $j$.  So, for example, if $j$ is a nearest-neighbor of site $i$, the right hand side of Eq.~(\ref{eq: inhomog_dyson}) is proportional to $t^2=t^{*2}/4d\rightarrow 0$ as $d\rightarrow\infty$.  Hence, we learn that $\delta G_{jj}(i\omega_n)=0$ for $j\ne i$ and large dimensions. If $G_{jj}$ is unchanged, then $\Sigma_j$ is
also unchanged. This means that
\begin{equation}
 \frac{\partial\Sigma_j(i\omega_n)}{\partial h_i}\propto \delta_{ij}.
\end{equation}
We now show how to derive one of the off-diagonal $c$-$f$ correlation functions.  We want to calculate
\begin{equation}
 \sum_{ij}t_{ij}\langle f^\dagger_if^{}_ic^\dagger_ic^{}_j\rangle=
\frac{1}{\beta}\sum_n\sum_{ij}\left [\frac{\partial}{\beta\partial h_i}+n_{fi}\right ]G_{ji}(i\omega_n),
\end{equation}
which follows directly from the definition of the operator average and an explicit computation of the derivative (the term multiplied by $n_{fi}$ arises from the derivative of the partition function). Now we focus on the derivative term, and use the $GG^{-1}G$ trick
\begin{eqnarray}
\frac{\partial}{\beta\partial h_i}G_{ji}(i\omega_n)&=&
\frac{\partial}{\beta\partial h_i}\sum_{kl}G_{jk}(i\omega_n)G^{-1}_{kl}(i\omega_n)G_{li}(i\omega_n)\\
&=&G_{ji}(i\omega_n)\left [ \frac{\partial}{\beta\partial h_i} \Sigma_i(i\omega_n)\right ] G_{ii}(i\omega_n),
\label{eq: h-deriv}
\end{eqnarray}
where we used the fact that the derivative of the self-energy was nonzero only for $k=l=i$. Since the self-energy is an implicit function of $G_{ii}$ and $n_{fi}$ one can compute the derivative of the self-energy with respect to the field by using the chain rule and re-expressing in terms of derivatives of the self-energy with respect to the Green function and the $f$-electron filling.  The algebra is quite long and is contained in Ref.~\onlinecite{Freericks3}.  The end result is that
\begin{equation}
 G_{ii}(i\omega_n)\left [ \frac{\partial}{\beta\partial h_i} \Sigma_i(i\omega_n)\right ]+n_{fi}=\frac{\Sigma_i(i\omega_n)}{U}.
\end{equation}
Plugging this result into Eq.~(\ref{eq: h-deriv}), and then converting the summation over $i$ and $j$ to a summation over momentum, produces Eq.~(\ref{corr_k_6}).

The only equation that requires some more formal development is Eq.~(\ref{corr_k_8}) because it involves two $f$-electron density operators, and hence derivatives with respect to two $h$ fields. Using the fictitious fields, one can immediately show that
\begin{eqnarray}
\sum_{ij}t_{ij}\langle
f_{j}^{\dagger}f^{}_{j}f_{i}^{\dagger}f^{}_{i}c_{i}^{\dagger}c^{}_{j}\rangle
=\sum_{ij}t_{ij}\left[\frac{\partial}{\beta\partial h_{j}}+n_{fj} \right]\left[\frac{\partial}{\beta\partial
h_{i}}+n_{fi} \right] G_{ji}(\tau =0^{-}). \label{A1}
\end{eqnarray}
All the terms in this expression, except the term proportional
$\partial^{2}G_{ji}/\partial h_{j}\partial h_{i}$, can be expressed
in terms of the Green function and self-energies by using the results above. In
order to find the second derivative of the Green function, one can show (similar
to the case of the first derivative) that in the limit of infinite
dimensions:
\begin{eqnarray}
\frac{\partial^{2}G_{ji}(i\omega_{n})}{\partial h_{j}\partial
h_{i}}|_{h=0}&=&\frac{\partial}{\partial h_{j}}\left[
G_{ii}(i\omega_{n})G_{ji}(i\omega_{n})\frac{\partial\Sigma_{i}}{\partial
h_{i}}\right]|_{h=0} \nonumber \\
&=&G_{ii}(i\omega_{n})G_{jj}(i\omega_{n})G_{ji}(i\omega_{n})\frac{\partial\Sigma_{j}}{\partial
h_{j}}\frac{\partial\Sigma_{i}}{\partial h_{i}}|_{h=0}
+G_{jj}(i\omega_{n})G_{ji}(i\omega_{n})\frac{\partial^{2}\Sigma_{i}}{\partial
h_{j}\partial h_{i}}|_{h=0}.\nonumber \\
\label{Gderiv3}
\end{eqnarray}
The last term in this equation is equal to zero, since the second
derivative of the self-energy
$\partial^{2}\Sigma_{i}(i\omega_{n})/\partial h_{j}\partial h_{i}$
vanishes. The argument is elementary.  Note that $\partial\Sigma_j/\partial h_j$ is a function of $G_{jj}$ and $\Sigma_j$. If we now take a derivative with respect to $h_i$ when $i\ne j$, the derivative must vanish, because the derivative of $G_{jj}$ and $\Sigma_j$ with respect to $h_i$ is zero. Therefore,
\begin{eqnarray}
\frac{\partial^{2}G_{ji}(i\omega_{n})}{\partial h_{j}\partial
h_{i}}|_{h=0}=G_{ii}(i\omega_{n})G_{jj}(i\omega_{n})G_{ji}(i\omega_{n})\frac{\partial\Sigma_{j}}{\partial
h_{j}}\frac{\partial\Sigma_{i}}{\partial h_{i}}|_{h=0}.
\label{Gderiv4}
\end{eqnarray}
Evaluating the derivatives explicitly and simplifying the final result then yields Eq.~(\ref{corr_k_8}).

\end{document}